\documentclass[prl,twocolumn,showpacs,superscriptaddress,preprintnumbers,amssymb]{revtex4}
\usepackage{graphicx}
\usepackage{dcolumn}
\usepackage{bm}
\usepackage{wasysym}
\input epsf

\newcommand{\beq}{\begin{equation}}

\begin{document}
\title{Doping dependence of thermopower and thermoelectricity in strongly correlated materials}
\author{Subroto Mukerjee}
\affiliation{Department of Physics, University of California, Berkeley, CA 94720}
\affiliation{Materials Sciences Division, Lawrence Berkeley National Laboratory, Berkeley, CA 94720}
\author{J.~E.~Moore}
\affiliation{Department of Physics, University of California, Berkeley, CA 94720}
\affiliation{Materials Sciences Division, Lawrence Berkeley National Laboratory, Berkeley, CA 94720}
\date{\today}

\begin{abstract}
The search for semiconductors with high thermoelectric figure of merit~\cite{goldsmid,rowe,disalvo} has been greatly aided by theoretical modeling of electron and phonon transport, both in bulk materials~\cite{mahan} and in nanocomposites~\cite{dresselhaus,gchen}.  Recent experiments~\cite{terasaki,wang,lee} have studied thermoelectric transport in  ``strongly correlated'' materials derived by doping Mott insulators, whose insulating behavior without doping results from electron-electron repulsion, rather than from band structure as in semiconductors.  Here a unified theory of electrical and thermal transport in the atomic and ``Heikes'' limit is applied to understand recent transport experiments on sodium cobaltate and other doped Mott insulators at room temperature and above. For optimal electron filling, a broad class of narrow-bandwidth correlated materials are shown to have power factors (the electronic portion of the thermoelectric figure of merit) as high at and above room temperature as in the best semiconductors.
\end{abstract}
\maketitle

The efficiency of a thermoelectric material for refrigeration or electrical generation is determined by its thermoelectric figure of merit

\begin{equation}
ZT = \frac{TS^2\sigma}{\kappa},
\end{equation}
where $T$ is the temperature, $\sigma$ is the electrical conductivity, $S$ the
thermopower and $\kappa$ the thermal conductivity. The same transport
coefficients that determine a material's utility for thermoelectric
applications are also of importance for fundamental materials science; for
example, thermopower measurements are frequently used to measure the sign of
the dominant charge carriers in a material. Most thermoelectrics currently in
use~\cite{disalvo} are bulk semiconductors such as Bi$_2$Te$_3$, with $ZT \approx 1$ at room temperature.

In recent years, strongly correlated oxides have been studied as candidate thermoelectric materials. Research in this direction was sparked off by the observation of high thermopower ($\sim 125 \mu V/K$) at room temperature coupled with a low electrical resistivity in the material Na$_x$CoO$_2$~\cite{terasaki,wang,lee}. Other materials including Sr$_x$La$_{1-x}$TiO$_3$ and several superconducting cuprates~\cite{okuda,obertelli} were found to have promising values of thermopower at room temperature. These materials are strongly correlated with an interaction energy scale $U$ that is much larger than the bandwidth set by the hopping $t$.  Theoretical calculations on such materials are difficult and often specific only to certain ranges of parameters~\cite{oguri,shastry}.

The  ``power factor'' $T S^2 \sigma$ appearing in the numerator of $ZT$ is
determined by properties of the charge carriers (electrons or holes), while
$\kappa$ is dominated by phonons for most viable thermoelectric materials.
Although applications of oxides as thermoelectric materials are currently
limited by their relatively high thermal conductivity, recent advances in
reducing phonon thermal conductivity in semiconductor
thermoelectrics~\cite{dresselhaus,gchen,venky} should lead to more dramatic
gains in oxides, where the original thermal conductivity is higher. Hence, understanding which oxides can have power factors as high or higher than the best semiconductors is an important step toward realizing the potential of this material class.

In this paper, we extend the calculation of the thermopower of correlated
systems in the atomic limit~\cite{beni,mukerjee} to the electrical and thermal
conductivities and obtain the power factor and electronic $ZT$, as functions
of doping and temperature.  Although the optimum band structure for
thermoelectric efficiency in a semiconductor was determined by Mahan and
Sofo~\cite{mahan}, the dominant role of electron-electron interactions in
correlated materials requires entirely different theoretical methods. The
power factor is found to display maxima at doping values of 5\% and 88 \% from
the Mott insulator consistent with recent experiments on
Na$_x$CoO$_2$~\cite{lee}. We discuss the effect of multiple orbitals in closing and comment on the search for high-efficiency thermoelectric oxides.

We describe a strongly correlated system by the Hubbard model, which consists
of a lattice with electrons capable of hopping from a site, with a single
orbital, to its nearest neighbors with an energy (hopping parameter)
$t$. Double ocuupancy incurs an energy cost of $U$ per site. This surprisingly
complex model has been used to study the physics of superconductivity, spin
and charge ordering, and the Mott transition seen in a variety of oxides. The
Mott insulator appears at the occupancy of one electron per site. At
temperature $T$, the atomic limit is defined as the approximation $t \ll
(k_BT, U)$. Systems with narrow bandwidths (e.g., Na$_x$CoO$_2$) are best
suited for this type of analysis. In this limit, $\sigma$, $S$ and $\kappa$ are given by the following expressions to leading order
\begin{equation}
\sigma = \frac{e^2A\beta}{Z_2}\left[e^{-\beta(U-3\mu)} + e^{\beta \mu}\right], \label{conductivity}
\end{equation}
\begin{equation}
S = -\frac{k_B}{e}\left[ \frac{\beta U e^{2\beta \mu}}{e^{\beta U} + e^{2\beta \mu}}-\beta \mu\right],\label{thermopower}
\end{equation}
and
\begin{equation}
\kappa = \frac{k_BA}{Z_2}\left[(\beta U)^2e^{-\beta(U-3\mu)}-\frac{(\beta U)^2e^{-2\beta(U-3\mu)}}{e^{-\beta(U-3\mu)+e^{\beta \mu}}} \right] +
\frac{\sigma t^2}{e^2T}. \label{thermal}
\end{equation}
Here $\beta = 1/k_BT$ and $A$ is a doping and temperature independent, but material dependent, factor given by
\begin{equation}
A=\frac{4\eta a^2t^2\tau}{V\hbar^2},
\end{equation}
where $a$ is the lattice spacing, $V$, the volume of the unit cell and $\tau$, the transport relaxation time. $\gamma$ is a geometry dependent
constant equal to $1$ for a square lattice and $2$ for a triangular lattice.
\begin{equation}
Z_2 = \left(1+2e^{\beta \mu}+e^{-\beta(U-2\mu)}\right)^2. \label{partition}
\end{equation}

The chemical potential $\mu$ is expressed in terms of the doping $\rho$ as
\begin{equation}
e^{\beta \mu} = \frac{\rho -1 + \sqrt{(\rho -1)^2 + \rho(2-\rho)e^{-\beta U}}}{(2-\rho)e^{-\beta U}}. \label{chemical}
\end{equation}
$\rho =1 $ is the Mott insulator and $\rho =0$ and $\rho = 2$ correspond to completely empty and full bands respectively and are both band
insulating limits. The second term in the expression for the thermal conductivity $\kappa$ is of sub-leading order at high temperature ($k_BT
\sim U$) but not in the Heikes limit $k_BT \ll U$, where it is the leading term. It should be noted that both $\sigma$ and $\kappa$ depend on a
relaxation time $\tau$ as in Drude theory, while $S$ and $ZT$ do not. There are further simplifications in the Heikes limit $k_BT \ll U$:
\begin{equation}
S = -\frac{k_B}{e} \log\left[\frac{2(1-\rho)}{\rho}\right],
\end{equation}
\begin{equation}
\sigma = \frac{e^2A\beta}{2}\rho(1-\rho),
\end{equation}
\begin{equation}
\kappa = \frac{\sigma t^2}{e^2T},
\end{equation}
and
\begin{equation}
ZT = \frac{e^2S^2}{k_B^2(\beta t)^2}
\end{equation}

The above expressions are for $\rho \leq 1$ and can be extended to $\rho \geq 1$ by making the substitution $\rho \rightarrow 2-\rho$. We note
that the thermopower has no temperature dependence and is independent of material parameters in this limit; it changes sign by going through a
resonance at the Mott insulator $\rho=1$.  Experimental data on Na$_x$CoO$_2$ is
compared with theory in Fig. 1. The electronic $ZT$ is divergent close to the
Mott insulator but adding a phonon contribution to $\kappa$ keeps it
finite. This is illustrated with the material parameters for Na$_x$CoO$_2$~\cite{kumar}, as an example in Fig. 2
\begin{figure}[h!]
\epsfxsize = 3.0in
\epsffile{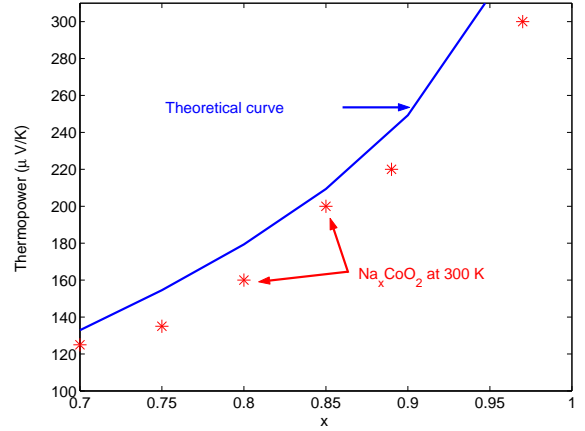}
\caption{The experimentally measured thermopower (asterisks) at different values of doping for
Na$_x$CoO$_2$ at 300 K. The theoretical Heikes curve is provided alongside for comparison. It can be seen that the experimental values are close
to the theoretical ones, showing that the Heikes limit is a good approximation
at 300 K. The values are systematically lower since the
Heikes limit is approached from below in temperature.}
\end{figure}

\begin{figure}[h!]
\epsfxsize = 3.0in
\epsffile{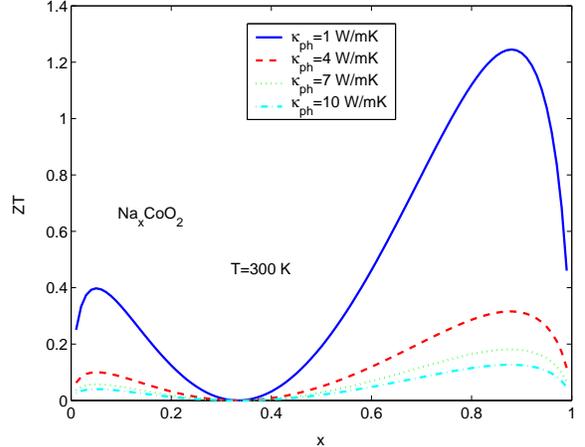}
\caption{$ZT$ at 300 K for different values of doping for Na$_x$CoO$_2$
  (with $t=110 K$) assuming different values of the the phonon thermal
  conductivity $\kappa_{ph}$ (assumed to be doping independent). $\kappa_{ph}$ adds to $\kappa$ from Eq. 10.} 
\end{figure}

The relaxation time $\tau = 2 \times 10^{-15}$ s for Na$_x$CoO$_2$ is extracted
from the conductivity measurements~\cite{lee}.  $x=0$
corresponds to the Mott insulator $\rho=1$. A plot of the theoretical
$TS^2\sigma$ for Na$_x$CoO$_2$ is shown in Fig. 3. The value of $TS^2\sigma$
($\approx 1-2 W{\rm m}^{-1}{\rm K}^{-1}$ is indeed observed in experiments. The main result of the above calculation is a robust
maximum in the power factor at dopings of $\rho =0.12$ and $\rho=1.88$ and secondary maxima close to the Mott insulator at
$\rho=0.95$ and $\rho=1.05$. These maxima arise from competition between the
(large) thermopower and (small) conductivity near the Mott ($\rho=1$) and band
($\rho=0$ and $\rho=2$) insulators. $\rho=1.88$ corresponds to x=0.88 for Na$_x$CoO$_2$, where a peak in the power factor has
indeed been observed in recent experiments over a range of temperatures~\cite{lee}.  Although we have used Na$_x$CoO$_2$ to enable comparison with experimental data, the fact that the main features are reproduced in the Hubbard model shows that {\it any} material with roughly similar values of $t$, $U$, and $\tau$ will have the same features.

\begin{figure}[h!]
\epsfxsize = 3.0in 
\epsffile{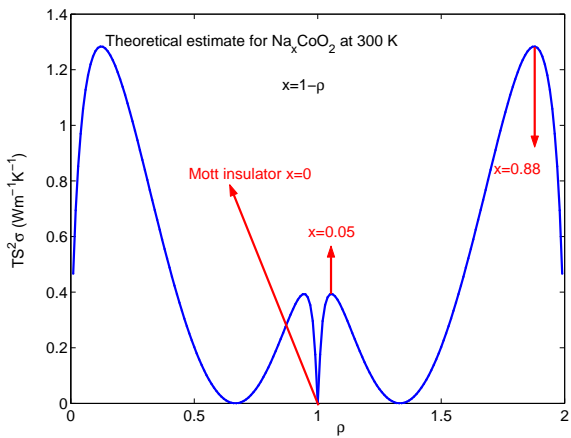}
\caption{The power factor ($TS^2\sigma$) as a function of doping for Na$_x$CoO$_2$ as predicted by
theory. There are maxima at $\rho \approx 0.12$, $\rho \approx 0.95$, $\rho \approx 1.05$ and $\rho \approx 1.88$. The last two values can be
realized in the material in principle at doping values $x=0.05$ and $x=0.88$. The positions of the peak are independent of the material
parameters in this calculation and a peak at $x=0.88$ has indeed been seen in experiments on Na$_x$CoO$_2$.}
\end{figure}

The location of the doping peak depends only weakly on temperature and for Na$_x$CoO$_2$ and shifts appreciably only at $T \geq 700$ K. The theoretical
position of the peak is independent of material dependent parameters. In reality however, factors like the effect of phonons on electron transport (phonon drag and polarons), orbital degrees of freedom, etc. could change the location of the peak.  As mentioned earlier, the above theory
initially assumed the atomic limit and hence is best applied to
narrow-bandwidth systems. For other correlated systems, one might need to consider terms beyond leading order in the calculation of transport
coefficients. However, the existence of a peak in the power factor is quite
general and has been observed in other candidate thermoelectric materials, such as
La$_x$Sr$_{1-x}$TiO$_3$ (for which polaronic effects are known to be important), close to the band insulator~\cite{okuda}.

We finally examine the effects of multiple orbitals per site on thermoelectric transport. This model is applicable to certain strongly
correlated materials like chromium and magnesium based oxides~\cite{maekawa}. We consider a multiple-orbital Hubbard model with $N$
degenerate orbitals per site and a single value of the hopping parameter (for
concreteness), and extend the calculation for the thermopower~\cite{mukerjee}
to include $\sigma$, $\kappa$, $ZT$ and $TS^2\sigma$.  There is a correlation energy
of $U$ for each pair of carriers occupying the same site. In the Heikes limit, we obtain the following expressions for the transport coefficients:

\begin{equation}
\sigma = \frac{e^2AN\beta}\rho(1-\rho),
\label{multicon}
\end{equation}
\begin{equation}
S = -\frac{k_B}{e} \log\left[\frac{N(1-\rho)}{\rho}\right],
\label{multis}
\end{equation}
\begin{equation}
\kappa = \frac{\sigma N^2 t^2}{e^2T},
\label{multitherm}
\end{equation}

$TS^2\sigma$ for an arbitrary system at a fixed temperature $T$ is shown in Fig. 4. There are still strong maxima as a function of doping, though
displaced slightly from the single-orbital ($N=1$) value. This is due mainly to the logarithmic dependence of $S$ on $N$. The overall magnitude increases for large $N$, as the presence of more hopping channels increases the conductivity.

\begin{figure}[h!]
\epsfxsize = 3.0in
\epsffile{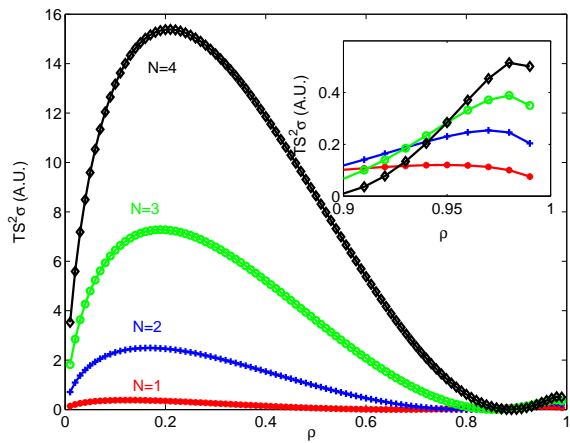}
\caption{The value of the power factor at a particular temperature for an
  arbitray system, as a function of doping for a multiple-orbital
Hubbard model. All material dependent parameters are assumed to be the same
for the different values of $N$. Larger $N$ implies more conducting
channels giving a larger conductivity and hence a larger value of $TS^2\sigma$. The peaks shift with increasing $N$ but not by much owing to a
weak logarithmic dependence of the thermopower on $N$.}
\end{figure}

To conclude, we have investigated the thermoelectric properties of strongly correlated systems by studying the Hubbard model. We have shown that
in the absence of phonon heat conduction, $ZT$ can diverge at values of doping
close to the Mott and band insulators. The calculated power factor and
thermopower agree with experiments on Na$_x$CoO$_2$. We have in particular
demonstrated the existence of a doping maximum 88 \% away from the Mott insulator, also been seen in experiments and have investigated the effect of multiple orbitals on it. The prediction of a large
power factor and the presence of a doping maximum should aid in searches for better thermoelectric oxides.

The authors wish to acknowledge conversations with R. Ramesh, A. Majumdar, C. Yu, and Y. Wang and support from DOE.

\bibliography{thermobib}

\end{document}